\documentclass{aa}
\usepackage{epsf}

\begin{document}
\title{$^1$S$_0$ neutron pairing vs observations of cooling neutron stars}
\author{
D. G. Yakovlev\inst{1}
\and
A. D. Kaminker \inst{1}
\and
O. Y. Gnedin\inst{2}
}
\institute{
Ioffe Physical Technical Institute,
         Politekhnicheskaya 26, 194021 St.~Petersburg, Russia \\
\and
Institute of Astronomy, Madingley Road, Cambridge CB3 0HA,
    England
\\
{\em  yak@astro.ioffe.rssi.ru,  kam@astro.ioffe.rssi.ru, 
 ognedin@ast.cam.ac.uk }}
\offprints{D.G.\ Yakovlev}

\date{Received 3 August 2001 / Accepted 24 September 2001}
\abstract{
As shown recently by Kaminker et al.\ (\cite{khy01}), 
current observations of thermal emission of isolated 
middle-aged neutron stars (NSs) can be explained
by cooling of NSs of different masses with the cores
composed of neutrons, protons and electrons, assuming
rather strong superfluidity (SF) of protons, weak triplet-state
SF of neutrons and neglecting singlet-state
SF of neutrons.
We show that this explanation remains correct in the
presence of singlet-state SF of neutrons 
in the NS crust and outermost core but under stringent
constraints on the density profile of the
SF critical temperature $T_{\rm cns}(\rho)$.
In order to explain observations
of (young and hot) RX J0822--43 and 
(old and warm) PSR 1055--52 and RX J1856--3754
as cooling not too massive 
NSs, the maximum $T_{\rm cns}^{\rm max}$ should be rather high
($\ga 5 \times 10^9$ K) 
and the decrease of $T_{\rm cns}(\rho)$
outside the maximum should
be sharp. These results place important constraints
on the models of nucleon SF in dense matter.
}
\titlerunning{$^1$S$_0$ pairing of neutrons 
vs observations of cooling neutron stars}
\authorrunning{D.G.\ Yakovlev et al.}
\maketitle

\section{Introduction}
\label{sect-intro}
Cooling of NSs depends on the poorly known properties of matter
in NS interiors. Combined with the best observational
data, the cooling theory 
can be used to 
explore the properties of this matter,
in particular,
critical temperatures $T_{\rm c}$ of SF 
of neutrons (n) and protons (p) (supplementing
very model-dependent microscopic calculations,
e.g., Lombardo \& Schulze \cite{ls01}).
SF of nucleons reduces the emissivity of neutrino
reactions and affects the nucleon heat capacity 
(e.g., Yakovlev et al.\ \cite{yls99}).
Moreover, SF initiates a specific 
neutrino emission associated with Cooper pairing of nucleons
(Flowers et al.\ \cite{frs76}).
In this way nucleon SF becomes a strong regulator of NS cooling.

Recently Kaminker et al.\ (\cite{khy01}, hereafter Paper I)
proposed the interpretation of observations of thermal
emission from eight isolated middle-aged NSs 
(Fig.\ 1)
using simple
models of cooling NSs with superfluid cores (composed
of neutrons, protons and electrons).
Accordingly, one can consider three types of
nucleon SFs with density dependent
$T_{\rm c}(\rho)$:
singlet-state ($^1$S$_0$) neutron SF in the NS inner crust
and the outermost core
($T_{\rm c}=T_{\rm cns}$); triplet-state ($^3$P$_2$) neutron SF 
in the core ($T_{\rm cnt}$);
and $^1$S$_0$ proton SF in the core
($T_{\rm cp}$). For simplicity,
in Paper I $^1$S$_0$ neutron SF was neglected.
The main reason was that, 
being mainly located in a thin crust, 
this SF seemed
unimportant for middle-aged NSs.
In Paper I the observations were explained
with the models of cooling NSs of different masses
assuming rather
strong proton SF and weak $^3$P$_2$
neutron SF. The
models require neither exotic composition
of NS core nor additional NS reheating.

We extend the results of Paper I
by including $^1$S$_0$ neutron superfluidity
and show that, contrary to our expectation, this
superfluidity is important.

\begin{figure}
\centering
\epsfxsize=86mm
\epsffile[20 143 590 720]{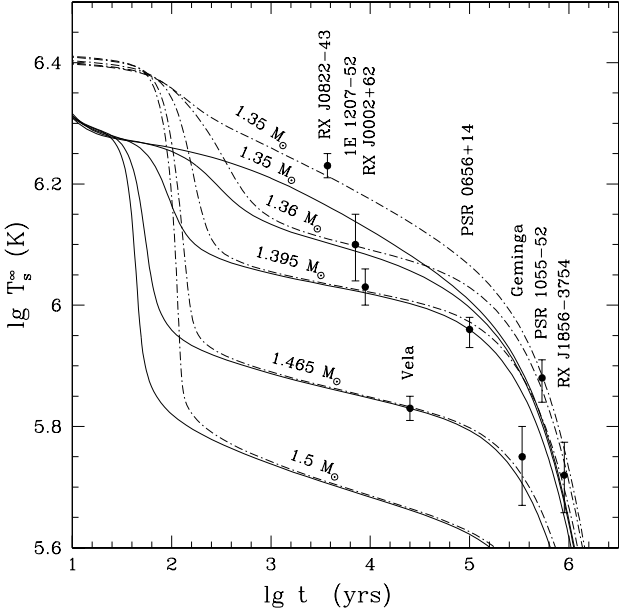}
\caption{
Observational limits on surface temperatures of eight
NSs compared with cooling curves
for NS models with masses from
1.35 to 1.5 ${\rm M}_\odot$.
Dot-and-dash curves are obtained
including proton SF in the NS core shown in Fig.\ 2.
Solid curves include, in addition,
model 1 of $^1$S$_0$ neutron SF from Fig.\ 2.
}
\label{fig-cool}
\end{figure}
 
\section{Cooling models with $^1$S$_0$ neutron pairing}
\label{sect-model}

We simulate NS cooling with our fully relativistic
nonisothermal cooling code (e.g., Gnedin et al.\
\cite{gyp01}). The physics input is mostly the same
as in Paper I.   In particular,
we adopt a moderately stiff 
equation of state (EOS) of matter in the NS core
proposed by Prakash et al.\ (\cite{pal88}) (their model I
with the compression modulus of 
saturated nuclear matter  $K=240$ MeV). 
The outer heat blanketing NS envelope is assumed to be composed
of iron,
but the atmosphere may be made of hydrogen. 
The maximum mass of our NS models is $
1.977\, {\rm M}_\odot$ 
(central density $
 \rho_{\rm c, max}=
2.575 \times 10^{15}$ g cm$^{-3}$).
For the given EOS, a powerful direct Urca process
of neutrino emission (Lattimer et al.\ \cite{lpph91})
is forbidden if
$M < M_{\rm D}=1.358 \, {\rm M}_\odot$ ($\rho_{\rm c} < 
7.851 \times 10^{14}$ g cm$^{-3}$).  In this case, a NS 
undergoes a {\it slow} cooling mainly via modified Urca process. 
A NS with $M>M_{\rm D}$ possesses a central kernel,
where direct Urca process
is open, initiating {\it fast} cooling.

We will confront theoretical cooling curves with
the same observational data as in Paper I; the only
exclusion is RX J1856--3754, for which new data
have become available.  
Figure 1 shows the limits
on the effective surface temperature,
$T_{\rm s}^\infty$, as measured by a distant observer,
versus stellar age $t$ for eight NSs.
The youngest three are radio-quiet NSs
in supernova remnants; the oldest is also a radio-quiet NS;
the others are observed  
as radio pulsars. 
The values of $T_{\rm s}^\infty$
are taken from the following sources: 
RX J0822--43 --- Zavlin et al.\ (\cite{ztp99}),
1E 1207--52 --- Zavlin et al.\ (\cite{zpt98}),
RX J0002+62 --- Zavlin \& Pavlov (\cite{zp99}),
PSR 0833--45 (Vela) --- Pavlov et al.\ (\cite{pavlovetal01}),
PSR 0656+14 --- Possenti et al.\ (\cite{pmc96}),
PSR 0630+178 (Geminga) --- Halpern \& Wang (\cite{hw97}), 
PSR 1055--52 --- \"Ogelman (\cite{ogelman95}), and
RX J1856--3754 --- Pons et al.\ (\cite{ponsetal01}).
As in Paper I the values of $T_{\rm s}^\infty$ 
for the four youngest NSs are inferred from the observations
using hydrogen atmosphere
models while for other SNs (but RX J1856)
they are inferred using the blackbody model.
For RX J1856, we take $T_{\rm s}^\infty= 0.52 \pm 0.07$ MK,
as inferred using the analytic fit with 
the Si-ash atmosphere model, and
$t=9 \times 10^5$ K (Pons et al.\ \cite{ponsetal01}).
We assume that wide errorbar of $T_{\rm s}^\infty$
given by this model is in line with current
poor understanding of thermal
emission from RX J1856 (also wee Burwitz et al.\ \cite{burwitzetal01}). 
It is also possible 
(Pons et al.\ \cite{ponsetal01}, Burwitz et al.\ \cite{burwitzetal01})
that the source has a colder surface
(0.25 MK) with a hot spot. This case would be easier
for our interpretation (see below) than the case we adopt.
 
\begin{figure}
\centering
\epsfxsize=70mm
\epsffile[20 143 575 510]{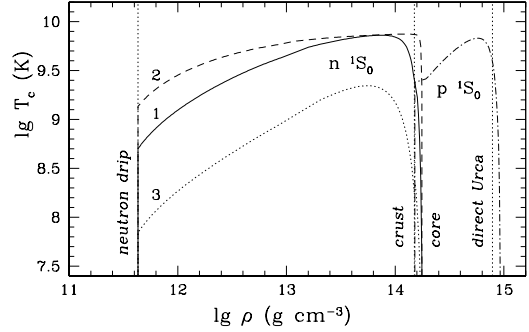}
\caption{
Density dependence of the critical temperatures
for one model of the proton SF (dots-and-dashes) and three models 
1, 2 and 3 of $^1$S$_0$ neutron SF (Table 1)
used in cooling
simulations (Figs.\ 1 and 3). Vertical
dotted lines indicate neutron drip,
core-crust interface, and the direct Urca threshold.
}
\label{fig-tc}
\end{figure}

We will not rely on any particular
(very model dependent)
microscopic calculation 
of SF critical temperature $T_{\rm c}(\rho)$. Instead,
following Paper I, we parameterize 
$T_{\rm c}$ as
\begin{equation}
    T_{\rm c}= T_0 \, { k^2 \over k^2 + k_1^2} 
     \; {(k-k_2)^2 \over  (k-k_2)^2 + k_3^2 }~,
\label{Tc}
\end{equation}
for $k < k_2$, and $T_{\rm c}=0$ for $k \geq k_2$.
Here, $k=k_{\rm F\!N}$
is the nucleon (N=n or p) Fermi wavenumber (measured in
fm$^{-1}$). 
The parameters 
$T_0$, $k_1$, $k_2$, $k_3$ 
of our SF models are given in Table 1.

As in Paper I we adopt a typical proton SF (Fig. 2).
We also assume that 
$^3$P$_2$ neutron SF in the NS core is weak
(maximum $T_{\rm cnt}^{\rm max} < 10^8$ K) and does not
affect cooling (see Paper I for details). 
Neglecting $^1$S$_0$ neutron SF,
we obtain the dot-and-dashed cooling curves displayed
in Fig.\ 1 for several values of $M$. The upper curve
is calculated for $M=1.35 \, {\rm M}_\odot$ but
it is actually the same for all $M$ from
$1.1 \, {\rm M}_\odot$ to $M_{\rm D}$.
The dot-and-dashed curves in Fig.\ 1 are slightly different from analogous
curves in Paper I since now we use a denser grid of
mass zones in the cooling code.
In agreement with Paper I we see that 
tuning $M$ we can explain the observations
of all eight sources.

Now let us include $^1$S$_0$ neutron SF.
Figure 2 shows the three models we consider.
Models 1 (solid line) 
and 2 (dashes)
correspond to about the same, rather strong SF
(with maximum $T_{\rm cns} \approx 7 \times 10^9$ K). 
The main difference is 
that $T_{\rm cns}(\rho)$ in model 2 has flatter maximum 
and sharper fall in the maximum wings. Model 3 (dots)
describes much weaker SF, with maximum 
$T_{\rm cns}\approx 2.4 \times 10^9$ K and a narrower density profile.   
 
The solid lines in Fig.\ 1 show cooling curves
calculated under the same assumptions as the dot-and-dashed lines
but including $^1$S$_0$ neutron SF
(model 1). As expected, the presence of this SF shortens the initial
thermal relaxation stage from 100 yrs to 30--50 yrs
(Gnedin et al.\ \cite{gyp01}) but does not
affect strongly the cooling of middle-aged NSs with open direct Urca
process ($M > M_{\rm D}$).  Thus,
$^1$S$_0$ neutron SF does not change
the results of Paper I for $M> M_{\rm D}$ and 
the proposed interpretation of 
observations of five sources,
1E 1207--52, RX J0002+62, Vela, PSR 0656+14, and Geminga.

\begin{table}[t]
\caption{Parameters of $^1$S$_0$ SF models in Eq.\ (1)}
\begin{tabular}{lllll}
SF & $T_{0}/10^9~$K & $k_1$, fm$^{-1}$  & $k_2$, fm$^{-1}$ & $k_3$, fm$^{-1}$\\
\hline
p  & 20.29     &   1.117  & 1.241   &  0.1473  \\
1n & 10.2      &   0.6    & 1.45    &  0.1     \\
2n & 7.9       &   0.3    & 1.45    &  0.01 \\
3n & 1800      &   21     & 1.45   &  0.4125 
\end{tabular}
\end{table}

However, this  SF drastically affects the
cooling of low-mass ($M \leq M_{\rm D}$) NSs.
Its main effects are to lower $T_{\rm s}^\infty$
at $t \la 3 \times 10^5$ yrs
by an additional neutrino emission associated with the
$^1$S$_0$ Cooper pairing of neutrons
and to lower $T_{\rm s}^\infty$ further at $t \ga 3 \times 10^5$ yrs
by reducing heat capacity of neutrons in the crust.
The importance of the Cooper emission 
can be explained as follows.
The neutrino emission from the core of a
middle-aged NS is strongly reduced by SF
suppression of the majority of neutrino reactions ($T \ll T_{\rm cp}$).
The neutrino emission due to $^1$S$_0$ neutron pairing 
is also greatly suppressed in the bulk of the inner NS
crust, where $T \ll T_{\rm cns}$, 
but remains active in two 
thin layers, where $T_{\rm cns}(\rho)$
is closer to the internal temperature $T$: 
near the neutron drip point 
and in the outermost part of the core.
Consider, for instance, the outermost core.
Since the emission layer is thin,
the emissivity can be written as
$Q^{\rm CP}=Q_0 T^7 F(T/T_{\rm cns})$,
where $Q_0$ may be regarded as constant (easily obtained
from Eq.\ 79 in Yakovlev et al.\ \cite{yls99}) 
and $F(\tau)$ is a known function. If we approximate
the dependence of $T_{\rm cns}$ on radial coordinate $r$
in our thin layer by a linear function
$T_{\rm cns}(r)=T+A(r-r_0)$ 
and integrate over this layer,
we obtain the Cooper--pairing (non-redshifted) neutrino 
luminosity:
%
\begin{equation}
  L^{\rm CP}=4 \zeta \pi r_0^2 Q_0 T^7 h, \quad
  \zeta =\int_0^1 {\rm d}\tau \,{F(\tau) \over \tau^2} =6.63,
\end{equation}
where $h=T/A$ is the characteristic
width of the layer (variation scale of $T_{\rm cns}(r)$ at
$r \approx r_0$). Adopting a typical value $r_0=12$ km
and the neutron number density $n_{\rm n}=0.1$ fm$^{-3}$,
we have
$L^{\rm CP} \approx 8.9 \times 10^{38} \, T_9^7 \, (h/100~{\rm m})$
erg s$^{-1}$, where $T_9 \equiv T/10^9$ K. For a typical scale
$h \sim 100$ m, the luminosity $L^{\rm CP}$ is comparable to
the standard (modified Urca) luminosity of the 
entire non-superfluid NS core! 
It may easily outweight the SF suppression
of the neutrino luminosity in other reactions.
In Paper I we analyzed analogous effect with regard
to $^3$P$_2$ neutron SF in the core and showed
that such SF, if available, strongly accelerates cooling
(even for $M > M_{\rm D}$)
and complicates interpretation of observations. 
Here we demonstrate that a similar but less pronounced
effect is produced by $^1$S$_0$ neutron SF at $M < M_{\rm D}$.

\begin{figure}
\centering
\epsfxsize=75mm
\epsffile[20 150 570 690]{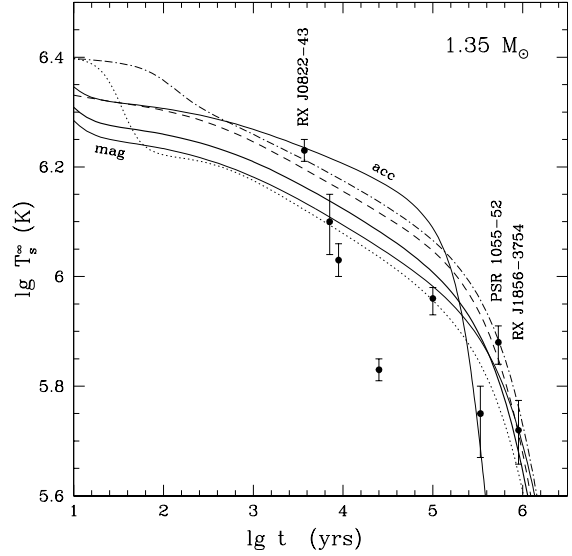}
\caption{
Cooling curves of the 1.35 ${\rm M}_\odot$ NS model
vs observations of RX J0822--43, PSR 1055--52
and RX J1856--3754. Dot-and-dashed curve: 
proton SF in the NS core. Solid, dashed
and dotted curves include, in addition,
models 1, 2 and 3 of $^1$S$_0$ neutron SF (Fig.\ 2), respectively.
Full solid line is the same as
in Fig.\ 1. Thin solid curve {\it acc} is calculated
assuming the presence of $10^{-10}\, {\rm M}_\odot$ of
hydrogen on the NS surface. Thin solid curve
{\it mag} is obtained assuming the dipole
surface magnetic field ($10^{12}$ G at the
magnetic pole).
}
\label{fig-coolh}
\end{figure}

According to Fig.\ 1, the $^1$S$_0$ neutron SF
complicates interpretation (Paper I) 
of observations of the youngest and hot source RX J0822--43 and
of two oldest and warm sources, PSR 1055--52 and RX J1856--3754
by the NS models with $M<M_{\rm D}$.
The less stringent complication for RX J1856--3754 
is evidently associated with too wide errorbar of $T_{\rm s}^\infty$. 
Let us focus (Fig.\ 3) on cooling of low--mass NS models
with regard to interpretation of these three sources.
For certainty, we again exploit the 1.35 ${\rm M}_\odot$
model although nearly the same is true
for any $M$ from $1.1 \, {\rm M}_\odot$ to $M_{\rm D}$.
To save the proposed interpretation
we must rise the upper cooling curve in Fig.\ 1.

The obvious solution is
to take $^1$S$_0$ neutron SF with steeper slopes
of $T_{\rm cns}(\rho)$ near the crust--core
interface and the neutron drip point. 
This decreases the characteristic
scale $h$
and the ``harmful'' luminosity $L^{\rm CP}$.
For example, taking model 2 of $^1$S$_0$ neutron SF
instead of model 1 (Fig.\ 2) we obtain the dashed
cooling curve (Fig.\ 3) which is closer
to the dot-and-dashed curve than the thick solid curve.
(Another example: shifting $T_{\rm cns}(\rho)$ for model
2 into the crust would additionally rise the cooling curves.) 
Note that the cooling curves
are insensitive to the details of the $T_{\rm cns}(\rho)$
profiles near the maximum of $T_{\rm cns}(\rho)$ as long as
$T_{\rm cns}^{\rm max}\ga 5 \times 10^9$ K,
but they are very sensitive to the decreasing
slopes of $T_{\rm cns}(\rho)$. 
A similar effect was reported in Paper I with regard to the
$T_{\rm cp}(\rho)$ profiles.
Taking smoother and lower 
$T_{\rm cns}(\rho)$, model 3, we obtain
a NS that is much colder than
required by observations (dotted line in Fig.\ 3).
Therefore,  $^1$S$_0$ neutron
SF with $T_{\rm cns}^{\rm max} < 5 \times 10^9$ K
and/or with smoothly decreasing slopes of the $T_{\rm cns}(\rho)$
profile near the crust-core interface and
neutron drip point {\it violates}
the proposed interpretation.    

Let us stress that the observations
of these three sources can be fit even with our
initial model 1 of $^1$S$_0$ neutron SF. The
high surface temperature of RX J0822--43 can be explained
assuming
the presence of low--mass
surface envelope
of light elements.  
This effect is modeled using 
the results of Potekhin et al.\ (\cite{pcy97}).   
Light elements increase the electron thermal conductivity
of NS surface layers and raise $T_{\rm s}^\infty$ at a
neutrino cooling stage, $t \la 3 \times 
10^5$ yrs (curve {\it acc} in Fig.\ 3).
In order to explain the observations of PSR 1055--52
and RX J1856--3754, we can assume again model 1 SF,
iron surface and
the dipole surface magnetic field
($\sim 10^{12}$ G at the magnetic pole;
line {\it mag}). 
Such a field makes the NS surface layers
overall less heat-transparent (Potekhin \& Yakovlev \cite{py01}), 
rising $T_{\rm s}^\infty$
at $t \ga 3 \times 10^5$ yrs. 
Note that the dipole field $\ga 10^{13}$ G
has the opposite effect,
similar to that produced 
by the surface envelope of light elements.
Thus, we can additionally vary cooling
curves by assuming the presence of light elements
and/or the magnetic field on the NS surface. However,
these variations are less pronounced than those produced by
nucleon SF. For instance, we cannot reconcile the cooling curves
with observations of PSR 1055--52
assuming model 3 of neutron SF
with any surface magnetic field.

\section{Summary}
\label{sect-sum}

Extending the results of Paper I, we have shown
that observations of thermal emission from eight
isolated NSs are consistent with
a simple model of cooling superfluid NSs
of different masses.
As in Paper I, the cooling model implies
rather strong proton SF and weak $^3$P$_2$
neutron SF in the NS core. In addition,
we have included the effects of $^1$S$_0$ neutron SF
in the crust and found that it
does not affect noticeably
interpretation of 1E 1207--52, RX J0002+62, Vela, PSR 0656+14, and
Geminga as NSs with masses $M>M_{\rm D}$.
However, this SF is crucial
for interpretation of RX J0822--43, PSR 1055--52,
and RX J1856--3754 (hotter for their ages) as NSs
with $M<M_{\rm D}$. 

The interpretation
requires the density profiles
of $T_{\rm cns}(\rho)$ 
with rather flat and not too low maximum ($\ga 5 \times 10^9$ K) and
steep decrease of $T_{\rm cns}(\rho)$ in the wings.
These requirements  
constrain the models of nn interaction in dense matter. 
We have compared the 
$^1$S$_0$ SF gaps $\Delta(k_{\rm Fn})=T_{\rm cns}/0.57$ 
for our models 1--3
with the results of numerous microscopic calculations,
shown in Fig.\ 7 in Lombardo \& Schulze (\cite{ls01}).
The gaps for models 1 and 3 
are
typical 
for microscopic theories in which medium effects
reduce $^1$S$_0$ pairing not too strongly. The gap
for model 2 is less ordinary (has too sharp decreasing
slope at large $k_{\rm Fn}$). 

Note, that our results 
depend crucially on the choice of $T_{\rm cp}(\rho)$
at $\rho \sim \rho_{\rm D}$. By taking
$T_{\rm cp}(\rho)$ shown in Fig.\ 2, we obtain
the NS masses in the range from $1.1 \, {\rm M}_\odot$ to
$1.47\, {\rm M}_\odot$, in a good agreement
with the well-known masses of radio pulsars
in binary systems (Thorsett \& Chakrabarty \cite{tc99}).
Adopting another $T_{\rm cp}(\rho)$ profile with
smoother decrease of $T_{\rm cp}$ at high density,
we would obtain higher masses for the same sources.
The inferred mass range is also affected by the choice of
EOS in the NS core (Paper I).
However, all these features
do not affect our conclusions on the
properties of $^1$S$_0$ neutron SF needed 
for the NS models
with $M < M_{\rm D}$.

One can see that the cooling of NSs
with $M < M_{\rm D}$ is sensitive to the density profile
of free neutrons near the crust bottom and
neutron drip point. We have used only one
model of free-neutron distribution in the crust, assuming
atomic nuclei to be spherical.
It would be interesting to consider the models
with non-spherical nuclei at the crust bottom
(e.g., Pethick \& Ravenhall \cite{pr95})
and take into account
SF of nucleons confined in atomic nuclei.

Let us stress that inferring $T_{\rm s}^\infty$
from observational data is the most complicated
problem (as discussed partly in Yakovlev et al.\ \cite{yls99}).
It would be important to improve the
limits on $T_{\rm s}^\infty$ and the
NS ages in the future high--quality observations;
they may change considerably.
Our interpretation is most sensitive to the data
on RX J0822--43, and especially on PSR 1055--52, and RX J1856--3754.
These sources seem
to be excellent for testing the cooling theories. 

\begin{acknowledgements}
We are grateful to the referee, G.G.\ Pavlov, for
useful suggestions, as well as to
K.\ Levenfish and A.\ Potekhin for critical remarks.
The work was supported partly by RFBR (grant No.\ 99-02-18099).
\end{acknowledgements}


\begin{thebibliography}{} 

\bibitem[2001]{burwitzetal01}
Burwitz, V., Zavlin, V.~E., Neuh\"auser, R., Predehl, R.,
Tr\"umper, J., \& Brinkman, A.~C. 2001, 
A\&A Lett.\ (accepted, astro-ph/0109374) 

\bibitem[2001]{gyp01}
Gnedin, O.~Y., Yakovlev, D.~G., \& Potekhin, A.~Y. 2001,
MNRAS 324, 725 

\bibitem[1976]{frs76}
Flowers, E.~G., Ruderman, M., \& Sutherland, P.G. 1976,
ApJ 205, 541

\bibitem[1997]{hw97}
Halpern, J.~P., \& Wang F.~Y.-H.
1997,
ApJ 477, 905

\bibitem[2001]{khy01}
Kaminker, A.~D., Haensel, P., \& Yakovlev, D.~G. 2001,
A\&A 373, L17 (Paper I) 

\bibitem[1991]{lpph91}
Lattimer, J.~M., Pethick, C.~J.,  Prakash, M., \& Haensel, P. 1991,
Phys.\ Rev.\ Lett.\  66, 2701

\bibitem[2001]{ls01}
Lombardo, U., \& Schulze, H.-J. 2001,
in Physics of Neutron Star Interiors,
eds.\ D.\ Blaschke, N.\ Glendenning, A.\ Sedrakian
(Springer, Berlin), p. 31 

\bibitem[1995]{ogelman95}
\"{O}gelman, H.
1995,
in Lives of Neutron Stars,
eds.\ M.~A.\ Alpar, \"U.\ Kizilo\u{g}lu, J.\ van Paradjis,
NATO ASI Ser.\
(Kluwer, Dordrecht) p.\ 101

\bibitem[2001]{pavlovetal01}
Pavlov, G.~G., Zavlin, V.~E., Sanwal, D., Burwitz, V., \& Garmire, G.~P.
2001,
ApJ, 552, L129

\bibitem[1995]{pr95}
Pethick, C.~J., \& Ravenhall, D.~G. 1995,
Ann.\ Rev.\ Nucl.\ Particle Sci.\ 45, 429

\bibitem[2001]{ponsetal01}
Pons, J., Walter, F., Lattimer, J., Prakash, M.,
Neuh\"{a}user, R., \& An, P. 2001, ApJ, submitted (astro-ph/0107404)

\bibitem[1996]{pmc96}
Possenti, A., Mereghetti, S., \& Colpi, M.
1996,
A\&A 313, 565

\bibitem[1997]{pcy97}
Potekhin, A.~Y., Chabrier, G., \& Yakovlev, D.~G.
1997,
A\&A 323, 415

\bibitem[2001]{py01}
Potekhin, A.~Y., \& Yakovlev, G.~G. 2001,
A\&A 374, 213

\bibitem[1988]{pal88}
Prakash, M., Ainsworth, T.~L., \& Lattimer, J.~M.
1988,
Phys.\ Rev.\ Lett.\ 61, 2518

\bibitem[1999]{tc99}
Thorsett, S.~E., \& Chakrabarty D.
1999,
ApJ 512, 288

\bibitem[1999]{yls99}
Yakovlev, D.~G., Levenfish, K.~P., \& Shibanov, Yu.~A. 
1999,
Physics--Uspekhi 42, 737 (astro-ph/9906456)

\bibitem[1998]{zpt98}
Zavlin, V.~E., Pavlov, G.~G., \& Tr\"{u}mper, J.
1998,
A\&A 331, 821

\bibitem[1999]{zp99}
Zavlin, V.~E., \& Pavlov, G.~G.
1999,
private communication

\bibitem[1999]{ztp99}
Zavlin, V.~E., Tr\"{u}mper, J., \& Pavlov, G.~G.
1999,
ApJ 525, 959


\end{thebibliography}
\end{document}